\documentclass[aps,pre,twocolumn,groupedaddress]{revtex4-1}

\usepackage{graphicx}
\usepackage{bm}
\begin{document}
\title{Turbulent diffusion of chemically reacting flows: theory and numerical simulations}
\author{T. Elperin$^1$}
\email{elperin@bgu.ac.il}
\homepage{http://www.bgu.ac.il/me/staff/tov}
\author{N. Kleeorin$^{1}$}
\email{nat@bgu.ac.il}
\author{M. Liberman$^{2}$}
\email{misha.liberman@gmail.com}
\author{A. N. Lipatnikov$^3$}
\email{andrei.lipatnikov@chalmers.se}
\author{I. Rogachevskii$^{1}$}
\email{gary@bgu.ac.il}
\homepage{http://www.bgu.ac.il/~gary}
\author{R. Yu$^4$}
\email{rixin.yu@energy.lth.se}
\vspace{1mm}
\affiliation{
$^1$The Pearlstone Center for
Aeronautical Engineering Studies, Department of
Mechanical Engineering, Ben-Gurion University of
the Negev, P. O. Box 653, Beer-Sheva
84105, Israel \\
$^2$Nordita, KTH Royal Institute of Technology and Stockholm University,
Roslagstullsbacken 23, 10691 Stockholm, Sweden \\
$^3$Department of Applied Mechanics, Chalmers University of Technology,
G\"{o}teborg, 412 96, Sweden\\
$^4$Division of Fluid Mechanics, Lund University, Lund, 221 00, Sweden \\
}
\vspace{1mm}
\begin{abstract}
The theory of turbulent diffusion of chemically reacting gaseous admixtures developed previously (Phys. Rev. E {\bf 90}, 053001, 2014) is generalized for large yet
finite Reynolds numbers, and the dependence of turbulent diffusion coefficient versus two parameters, the Reynolds number and Damk\"ohler number (which characterizes a ratio of turbulent and reaction time scales) is obtained.
Three-dimensional direct numerical simulations (DNS) of a finite thickness reaction wave for the first-order chemical reactions propagating in forced, homogeneous, isotropic, and incompressible turbulence are performed to validate the theoretically predicted effect of chemical reactions on turbulent diffusion.
It is shown that the obtained DNS results are in a good agreement with the developed theory.
\end{abstract}
\date{\today}
\vspace{-10mm}

\maketitle

\section{Introduction}

Effect of chemical reactions on turbulent transport is of great importance in many applications ranging from atmospheric turbulence and transport of pollutants to combustion processes (see, e.g., \cite{CSA80,BLA97,F03,P04,SP06,ZA08,ML08,CRC}).
For instance, significant influence of combustion on turbulent transport is well known \cite{CRC,L85,Br95,PECS10,SB11,AR17} to cause
the so-called counter-gradient scalar transport, i.e. a flux of products from unburnt
to burnt regions of a premixed flame.
In its turn, the counter-gradient transport can substantially reduce the flame speed \cite{SL13,SL14,SL15} and, therefore,
is of great importance for calculations of
burning rate and plays a key role in the premixed turbulent combustion.

It is worth remembering, however, that
the counter-gradient transport appears to be an indirect manifestation of the influence of chemical reactions on turbulent fluxes,
as this manifestation is controlled by density variations due to heat release in combustion reactions, rather than by the reactions themselves.
As far as the straightforward influence of reactions on turbulent transport \cite{Corr} is concerned, such effects
have yet been addressed in a few studies \cite{CRC,BD17,L33} of premixed flames.
Because the easiest way to studying such a straightforward influence consists in investigating a constant-density reacting flow,
the density is considered to be constant in the present paper.

The effect of chemical reactions on turbulent diffusion
of chemically reacting gaseous admixtures in a developed turbulence
has been studied analytically using a path-integral approach for a
delta-correlated in time random velocity field \cite{EKR98}.
This phenomenon also has been recently investigated
applying the spectral-tau approach that is valid for
large Reynolds and Peclet numbers \cite{EKLR14}.
These studies have demonstrated that turbulent diffusion of the reacting species
can be strongly suppressed with increasing
Damk\"{o}hler number, ${\rm Da}=\tau_0/\tau_c$,
that is a ratio of turbulent, $\tau_0$, and chemical, $\tau_c$, time scales.

The dependence of turbulent diffusion coefficient, $D_T$, versus
the turbulent Damk\"{o}hler number obtained theoretically
in \cite{EKLR14}, was validated using results of mean-field simulations
(MFS) of a reactive front propagating in a turbulent flow \cite{BH11}.
In these simulations, the mean speed, $s_{_T}$, of the
planar one-dimensional reactive front was determined
using numerical solution of the
Kolmogorov-Petrovskii-Piskunov (KPP)
equation \cite{KPP37} or the Fisher equation \cite{F37}.
This mean-field equation was extended in \cite{BH11} to take into account
memory effects of turbulent diffusion
when the turbulent time was much larger than the characteristic
chemical time.
Turbulent diffusion coefficients as a function of ${\rm Da}$
were determined numerically in \cite{BH11} using the obtained function $s_{_T}({\rm Da})$
and invoking the well-known expression, $s_{_T}=2(D_T/\tau_c)^{1/2}$.
The theoretical dependence $D_T({\rm Da})$ derived in \cite{EKLR14}
was in a good agreement with the numerical results of MFS \cite{BH11}.

In the present study we have generalized the theory \cite{EKLR14}
of turbulent diffusion in reacting flows for
finite Reynolds numbers and have obtained the dependence of turbulent
diffusion coefficient versus two parameters, the Reynolds number
and Damk\"ohler number.
The generalized theory has been validated
by comparing its predictions with the three-dimensional direct
numerical simulations (DNS) of the reaction wave propagating in
a homogeneous isotropic and incompressible turbulence
for a wide range of ratios of the wave speed to
the r.m.s. turbulent velocity and different Reynolds numbers.

It is worth stressing that the previous validation of the original theory
by MFS \cite{EKLR14} and present validation of the generalized theory
by DNS complement each other,
because they were performed using different methods.
Indeed, the previous validation \cite{EKLR14} was performed
by evaluating $D_T$ using numerical data \cite{BH11} on the mean
reaction front speed obtained by solving a statistically planar one-dimensional
mean KPP equation, with such a MFS method
implying spatial uniformity of the turbulent diffusion coefficient.
On the contrary, in the present work, $D_T$ is straightforwardly extracted from
DNS data obtained by numerically integrating
unsteady three-dimensional Navier-Stokes and reaction-diffusion
equations, with eventual spatial variations in the turbulent
diffusion coefficient being addressed.

This paper is organised as follows.
The generalized theory is described in Section~II.
DNS performed to validate the theory are described in Section~III.
Subsequently, validation results are discussed in Section~IV,
and concluding remarks are given in Section~V.

\section{Effect of chemistry on turbulent diffusion}

The goal of this section is to generalize the theory \cite{EKLR14} by considering turbulent flows characterized by large, but finite Reynolds numbers.
It is worth stressing that neither the original theory \cite{EKLR14} nor its generalization have specially been developed to study combustion.
On the contrary, while the theory addresses a wide class of turbulent reacting flows, certain
assumptions of the theory do not hold in premixed flames.
Nevertheless, as will be shown in subsequent sections, the theoretical
predictions are valid under a wider range of conditions than originally assumed and, in particular, under conditions associated with
the straightforward influence of chemical reactions on turbulence in flames.

\subsection{Governing equations}

Equation for the scalar field in the incompressible
chemically reacting turbulent flow reads:
\begin{eqnarray}
\frac{\partial c}{\partial t} + ({\bm v} \cdot
{\bm  \nabla}) c  = W(c) + D \Delta c,
 \label{A1}
\end{eqnarray}
where $c(t,{\bm x})$  is a scalar field,
${\bm v}(t,{\bm x})$ is the instantaneous
fluid velocity field,
$D$ is a constant diffusion coefficient
based on molecular Fick's law,
$W(c)$ is the source (or sink) term.
The function $W(c)$ is usually chosen according to the Arrhenius law
(to be given in the next section).
We consider a simplified model of a single-step reaction
typically used in numerical simulations of turbulent combustion.

The velocity ${\bm v}$ of the fluid is determined by the Navier-Stokes equation:
\begin{eqnarray}
{\partial {\bm v} \over \partial t} &+& ({\bm v}
\cdot {\bm  \nabla}) {\bm v} = - {1 \over \rho}
{\bm  \nabla} p + \nu \Delta {\bm v} + {\bm f},
 \label{A2}
\end{eqnarray}
where ${\bm f}$ is the external force to support turbulence,
$\nu$ is the kinematic viscosity,
$p$ and $\rho$ are the fluid pressure and density, respectively.
For an incompressible flow the fluid density is constant.

\subsection{Procedure of derivations of turbulent flux}

To determine turbulent transport coefficients, Eq.~(\ref{A1})
is averaged over an ensemble of
turbulent velocity fields. In the framework of a mean-field
approach, the scalar field $c$  is decomposed into the
mean field, $\langle c \rangle$, and fluctuations, $c'$,
where $\langle c' \rangle=0$,
and angular brackets imply
the averaging over the statistics of turbulent
velocity field. The velocity
field is decomposed in a similar fashion,
${\bm v}=\langle {\bm U} \rangle +{\bm u}$, assuming for
simplicity vanishing mean fluid velocity, $\langle {\bm U} \rangle=0$,
where ${\bm u}$ are the fluid velocity
fluctuations.

Using the equation for fluctuations $c'=c-\langle c \rangle$ of the
scalar field and the Navier-Stokes equation
for fluctuations ${\bm u}$ of the velocity field written in ${\bm k}$ space we derive
equation for the second-order moment $\langle c' \,{\bm u} \rangle_{\bm k}
\equiv\langle c'({\bm k}) \,u_i(-{\bm k})
\rangle$:
\begin{eqnarray}
&&\frac{\partial \langle c' \,u_i
\rangle_{\bm k}}{\partial t} = - \left[\tau_c^{-1} + (\nu + D) k^2\right]
\langle c' \,u_i \rangle_{\bm k} + \hat {\cal N} \langle c'
\,u_i \rangle
\nonumber\\
&&\quad - \langle u_i \,u_j \rangle_{\bm k} \, \nabla_j
\langle c \rangle ,
 \label{C5}
\end{eqnarray}
where $\langle u_i \,u_j \rangle_{\bm k}
\equiv \langle u_i({\bm k}) \,u_j(-{\bm k})
\rangle$, $\tau_c= \langle c \rangle /\langle W \rangle$
is the chemical time,
$\langle W \rangle$ is the mean source function
and $\hat {\cal N} \langle c' \,u_i \rangle$
includes the third-order moments caused by the
nonlinear terms:
\begin{eqnarray}
&& \hat {\cal N} \langle c' \,u_i \rangle = -
\langle [{\bm \nabla \cdot} (c' \,{\bm u})]
\,u_i \rangle_{\bm k} - \langle c' \, [({\bm u}\cdot {\bm
\nabla}) u_i] \rangle_{\bm k}
\nonumber\\
&& \quad -\langle c' \, [\rho^{-1} \nabla_i p'] \rangle_{\bm k} .
 \label{CC5}
\end{eqnarray}
Here we follow the procedure of the derivation of the turbulent fluxes
that is described in detail in \cite{EKLR14},
taking into account large yet finite Reynolds number.
In particular, we use multi-scales approach
(i.e., we separated fast and slow variables, where fast small-scale variables correspond to
fluctuations and slow large-scale variables correspond to
mean fields).
Since the ratio of spatial density of species is assumed to be
much smaller than the density of the surrounding fluid (i.e., small mass-loading
parameter), there is only one-way coupling, i.e.,
no effect of species on the fluid flow. Due to the same reason the energy release
(or absorbtion of energy) caused by chemical reactions is
much smaller than the internal energy of the surrounding
fluid. This implies that even small chemical time does
not affect the fluid characteristics.
Finally, we also assume that the deviations of the source term $W$
from its mean value $\langle W \rangle$ is not large.
While such an assumption does not hold in a typical premixed turbulent
flame \cite{CRC}, we will see later that the theory
well predicts the effect of the chemical reaction on turbulent transport even if difference in $W(c)$ and $\langle W \rangle(\langle c \rangle)$
is significant, as commonly occurs in the case of premixed combustion.

The equation for the second-order
moment~(\ref{C5}) includes the first-order
spatial differential operators applied to the
third-order moments $\hat {\cal N} \langle c' \,u_i \rangle$.
To close the system of equations it is necessary to express
the third-order terms $\hat {\cal N} \langle c' \,u_i \rangle$
through the lower-order moments $\langle c' \,u_i \rangle_{\bm k}$ (see,
e.g., \cite{O70,MY75,Mc90}). We use the spectral
$\tau$ approximation that postulates that the
deviations of the third-order moments,
$\hat {\cal N} \langle c' \,u_i \rangle$, from the contributions to
these terms afforded by the background
turbulence, $\hat {\cal N} \langle c' \,u_i \rangle^{(0)}$,
can be expressed through the similar deviations
of the second-order moments, $\langle c' \,u_i \rangle_{\bm k} -
\langle c' \,u_i \rangle_{\bm k}^{(0)}$:
\begin{eqnarray}
&& \hat {\cal N} \langle c' \,u_i \rangle - \hat {\cal N} \langle c' \,u_i \rangle^{(0)}
= - {1 \over \tau_r(k)} \,
\Big[\langle c' \,u_i \rangle_{\bm k} - \langle c' \,u_i \rangle_{\bm k}^{(0)}\Big],
\nonumber\\
 \label{D1}
\end{eqnarray}
(see, e.g., \cite{O70,MY75,PFL76}), where
$\tau_r(k)$ is the scale-dependent relaxation
time, which can be identified with the
correlation time $\tau(k)$ of the turbulent
velocity field for large Reynolds and Peclet
numbers. The functions with the superscript $(0)$
correspond to the background turbulence with zero
gradients of the mean scalar field.
Validation of the $\tau$ approximation
for different situations has been performed in
various numerical simulations and analytical
studies (see, e.g., \cite{BS05,RK07,RKKB11}).
When the gradients of the mean
scalar field are zero, the turbulent flux vanishes, and the contributions
of the corresponding fluctuations [the terms with
the superscript (0)], vanish as well.
Consequently, Eq.~(\ref{D1}) reduces to
$\hat{\cal N} \langle c' \,u_i\rangle_{\bm k} =
- \langle c'({\bm k}) \,u_i(-{\bm k}) \rangle /
\tau(k)$.

We also assume that the characteristic
times of variation of the second-order moments are
substantially larger than the correlation time
$\tau(k)$ for all turbulence scales.
This allows us to consider the steady-state solution of Eq.~(\ref{C5}),
that yields the following expression for the turbulent flux $\langle c' \,u_i \rangle_{\bm k} =\langle c'({\bm k}) \,u_i(-{\bm k}) \rangle$ in ${\bm k}$ space:
\begin{eqnarray}
&&\langle c' \,u_i \rangle_{\bm k} = - \tau_{\rm eff}(k) \, \langle u_i \,u_j \rangle_{\bm k}^{(0)} \, \nabla_j \langle c \rangle ,
 \label{C6}
\end{eqnarray}
where $\tau_{\rm eff}(k) = \left[\tau_c^{-1} + (\nu + D) k^2 +\tau^{-1}(k)\right]^{-1}$
is the effective time.

We consider isotropic and homogeneous background turbulence, $\langle u_i \,u_j \rangle_{\bm k}^{(0)} \equiv \langle u_i({\bm k}) \, u_j(-{\bm k}) \rangle$ (see, e.g., \cite{B53}):
\begin{eqnarray}
\langle u_i({\bm k}) \, u_j(-{\bm k}) \rangle =
{u_0^2 \, E_T(k) \over 8 \pi k^2}
\Big[\delta_{ij} - {k_i \, k_j \over k^2} \Big],
 \label{D3}
\end{eqnarray}
where
\begin{eqnarray}
E_T(k)={2 \over 3k_{0}} \left(1- {\rm Re}^{-1/2}\right)^{-1} \left({k \over
k_{0}}\right)^{-5/3} ,
 \label{DD3}
\end{eqnarray}
is the energy spectrum function
for $k_0 \leq k \leq k_0 {\rm Re}^{3/4}$, $\tau(k) = 2 \,
\tau_0 \, (k / k_{0})^{-2/3}$ is the turbulent
correlation time, $k_{0}=\ell_0^{-1}$,
and ${\rm Re} = \ell_0 u_0/\nu \gg 1$ is the Reynolds number,
$u_0$ is the characteristic turbulent velocity in the integral scale,
$\ell_0$, of turbulence. For comparison of the theory with DNS we
do not neglect the small ${\rm Re}^{-1/2}$ term in Eq.~(\ref{DD3}).

\subsection{Turbulent flux}

After integration in ${\bm k}$ space we obtain
the expression for the turbulent flux, $\langle
c' \, {\bm u} \rangle$:
\begin{eqnarray}
\langle c' \, {\bm u} \rangle = \int \langle c' \,u_i \rangle_{\bm k} \,d{\bm k}
= - D_T \, {\bm \nabla} \langle c \rangle  ,
 \label{A11}
 \end{eqnarray}
where the coefficient of turbulent diffusion $D_T$ of the scalar field is
\begin{eqnarray}
D_T= {D^T_0 \over {\rm Da}} \, \left[1- {\Phi({\rm Da},{\rm Re},{\rm Pr}) \over 1- {\rm Re}^{-1/2}}  \right],
 \label{A12}
 \end{eqnarray}
$D^T_0 = \tau_0 u_0^2/3$ is the characteristic value of the turbulent
diffusion coefficient without chemical reactions,
$\tau_0=\ell_0/u_0$ is the characteristic turbulent time,
the function $\Phi({\rm Da},{\rm Re},{\rm Pr})$ is
\begin{eqnarray}
\Phi({\rm Da},{\rm Re},{\rm Pr}) = \int_{{\rm Re}^{-1/2}}^{1} \,  {X^2 + a({\rm Re},{\rm Pr}) \over 2 {\rm Da} X^3 + X^2 + a} \,dX ,
\label{A14}
 \end{eqnarray}
the parameter $a({\rm Re},{\rm Pr})=2\left(1+{\rm Pr}^{-1}\right)/{\rm Re}$, and ${\rm Pr}=\nu/D$ is the Prandtl number.

Evaluating approximately the integral in Eq.~(\ref{A14}) by expanding
the expression in the integral over small parameter $a({\rm Re},{\rm Pr})$
for large yet finite Reynolds numbers,
we obtain the following dependence of turbulent diffusion
coefficient versus Damk\"ohler and Reynolds numbers:
\begin{eqnarray}
D_T &=& {D^T_0 \over {\rm Da}} \, \left[1- {1 \over 2{\rm Da} \left[1- {\rm Re}^{-1/2}\right]} \, \ln {1 + 2{\rm Da} \over 1 + 2{\rm Da} \, {\rm Re}^{-1/2}} \right]
\nonumber\\
&& - 2 D^T_0 \, \left(1+{\rm Pr}^{-1}\right) \, {\ln {\rm Re} \over {\rm Re}} .
 \label{CA17}
\end{eqnarray}
In the limit of extremely large Reynolds numbers, we recover the result
for the function $D_T ({\rm Da})$ obtained in \cite{EKLR14}:
\begin{eqnarray}
D_T &=& {D^T_0 \over {\rm Da}} \, \left[1- {\ln (1 + 2{\rm Da})
\over 2{\rm Da}} \right].
\nonumber\\
 \label{A17}
\end{eqnarray}
It follows from  Eq.~(\ref{CA17}) that for small
Damk\"{o}hler numbers, ${\rm Da} \ll 1$,
the function $D_T ({\rm Da})$ is given by:
\begin{eqnarray}
D_T &=& D^T_0 \, \biggl[1- {4 {\rm Da} \over 3}  - 2 D^T_0 \, \left(1+{\rm Pr}^{-1}\right) \, {\ln {\rm Re} \over {\rm Re}}
\nonumber\\
&&+ {\rm Re}^{-1/2} \biggr],
 \label{CA18}
\end{eqnarray}
while for large Damk\"{o}hler numbers, $1 \ll {\rm Da} \ll {\rm Re}^{1/2}$, it is
\begin{eqnarray}
D_T={D^T_0 \over {\rm Da}} \, \left[1- {\ln 2{\rm Da} \over 2{\rm Da}}  - 2 \, \left(1+{\rm Pr}^{-1}\right) \, {{\rm Da} \, \ln {\rm Re} \over {\rm Re}}\right] ,
\nonumber\\
 \label{CA19}
\end{eqnarray}
and for very large Damk\"{o}hler numbers, $1 \ll {\rm Re}^{1/2} \ll {\rm Da}$,
it is
\begin{eqnarray}
D_T={D^T_0 \over {\rm Da}} \, \left[1- 2 \, \left(1+{\rm Pr}^{-1}\right) \, {{\rm Da} \, \ln {\rm Re} \over {\rm Re}} \right].
 \label{CA20}
\end{eqnarray}

Equations~(\ref{CA19}) and~(\ref{CA20}) show that turbulent
diffusion of particles or gaseous admixtures for a large
Damk\"{o}hler number, ${\rm Da} \gg 1$ is strongly reduced,
i.e., $D_T = D_0^T/{\rm Da} = \tau_{c} u_0^2/3$.
This implies that the turbulent diffusion for a large turbulent
Damk\"{o}hler number is determined by the chemical time.
The underlying physics of the strong reduction of turbulent
diffusion is quite transparent.
For a simple first-order chemical
reaction $A \to B$ the species $A$ of the
reactive admixture are consumed and their
concentration decreases much faster during the
chemical reaction, so that the usual turbulent
diffusion based on the turbulent time $\tau_0 \gg
\tau_c$, does not contribute to the mass flux of
a reagent $A$.

\section{DNS model}

Direct numerical simulations of a finite thickness reaction wave propagation
in forced, homogeneous, isotropic, and incompressible turbulence
for the first-order chemical reactions,
were performed in a fully periodic rectangular box of size
of $L_x \times L_y \times L_z$ using a uniform rectangular mesh of
$N_x \times N_y \times N_z$ points and a simplified in-house
solver \cite{YYB12} developed for low-Mach-number reacting flows.
Contrary to recent DNS studies by two of the present authors
\cite{YLB14,YBL15} that addressed self-propagation of an infinitely
thin interface by solving a level set equation, the present simulations
deal with a wave of a finite thickness, modelled with
Eq.~(\ref{A1}) for a single progress variable $c$ ($c=0$ and 1 in reactants and products, respectively), while the Navier-Stokes equation~(\ref{A2})
was numerically integrated in both cases.

To mimic a highly non-linear dependence of the reaction rate $W$
on the scalar field $c$ in a typical premixed flame characterized by significant variations
in the density $\rho$ and temperature $T$, the following expression,
\begin{eqnarray}
W= {1-c \over \tau_{_R} \, (1+ \tau)} \exp \left[-{{\rm Ze}\, (1+ \tau)^2 \over \tau \, (1+ c \, \tau)}  \right] ,
 \label{A18}
\end{eqnarray}
was invoked in the present constant-density simulations.
Here, $\tau_{_R}$ is a reaction time scale,
while parameters $Ze=6.0$ and $\tau=6.0$ are counterparts of the Zeldovich number ${\rm Ze} =E_a (T_b-T_u)/R T_b^2$ and
heat-release factor $(\rho_u-\rho_b)/\rho_b$, respectively, which are widely used in the combustion theory \cite{CRC,L85,Br95},
with subscripts $u$ and $b$ designating unburned and burned mixtures, respectively.
Indeed, substitution of $c=(T-T_u)/(T_b-T_u)$ into exponent in Eq.~(\ref{A18}) results in the classical Arrhenius law
\begin{eqnarray}
W= {1-c \over \tau_{_R} \, (1+ \tau)} \exp \left(-\frac{E_a}{R T} \right),
 \label{A19}
\end{eqnarray}
i.e. Eq.~(\ref{A18}) does allow us to mimic behavior of reaction rate in a flame by considering constant-density reacting flows.
It is worth remembering that a simplification of a constant density is helpful for studying the straightforward influence of chemical
reactions on turbulent transport, as already pointed out in Sect. I.

The speed $S_{_L}$ of the propagation of the reaction wave in the laminar flow, the wave thickness $\delta_{_F}=D/S_{_L}$, the wave time scale $\tau_{_F}=\delta_{_F}/S_{_L}$
were varied by changing the diffusion coefficient $D$ and the reaction time scale $\tau_{_R}$, which were constant input parameters for each DNS run.
The speed $S_{_L}$ was determined by numerically solving one-dimentional
Eq. (\ref{A1}) with ${\bm v}=0$.

The present DNS are similar to DNS discussed
in detail in \cite{YLB14,YBL15},
except for substitution of a level set equations used in \cite{YLB14,YBL15}
by Eqs.~(\ref{A1}) and (\ref{A18}).
Therefore, we will restrict ourselves to a very brief summary
of the simulations.
A more detailed discussion of the simulations can be found in recent papers \cite{PRE17RA,PoF17RA}.

The boundary conditions were periodic not only in transverse directions $y$ and $z$, but also in direction $x$ normal to the mean
wave surface. In other words, when the reaction wave reached the left boundary ($x=0$) of the computational domain,
the identical reaction wave entered the domain through its right boundary ($x=L_x$).

\begin{figure}
\vspace*{1mm} \centering
\includegraphics[width=8.0cm]{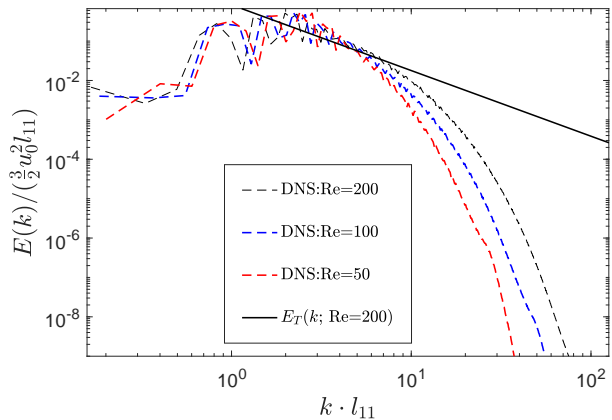}
\caption{\label{Fig0} Spectrum of turbulent kinetic energy}
\end{figure}

The initial turbulence field was generated by synthesizing prescribed Fourier waves \citep{YB14} with an initial rms velocity $u_0$
and the forcing scale
$\ell_{\rm f}=L/4$, where $L=L_y=L_z=L_x/4$ is the width of the computational domain.
Subsequently, a forcing function ${\bm f}$, see Eq.~(\ref{A2}), was invoked to maintain statically stationary turbulence following
the method described in Ref.~\cite{GLMA95}.
As shown earlier \cite{YLB14,YBL15},
(i) the rms velocity $u_0$ was maintained as the initial value,
(ii) the normalized dissipation rate
$\langle \varepsilon \rangle \ell_{\rm f}/u_0^3$
averaged over the computational domain fluctuated slightly above 3/2 after a short period
($t<\tau_{\rm f}=\ell_{\rm f}/u_0$)
of rapid transition from the initial artificially synthesized flow
to the fully developed turbulence,
(iii) the forced turbulence achieved good statistical homogeneity and isotropy over the entire domain, and
(iv) the energy spectrum showed a sufficiently wide range of the Kolmogorov scaling ($-5/3$) at the Reynolds number, ${\rm Re}\equiv u_0 \ell_{\rm f}/\nu=200$, based
on the scale $\ell_{\rm f}$
(see Fig.~\ref{Fig0}).

\begin{table}[t]
        \centering
        \caption{DNS cases.}
        \begin{tabular}{ccccccc}
\hline
\hline
Case &\,  ${\rm Re}$& \, ${\rm Re}_{\lambda}$ & \, $\eta/\Delta x$ & \, $S_{_L}/u_0$ & \, $\ell_{11}/\delta_{_F}$ & \, ${\rm Da}_{_{\rm DNS}}$ \\
\hline
                1& 50      & 18 & 0.68 & 0.1 & 2.1       & 0.2 \\
                2& 50      & 18 & 0.68 & 0.2 & 2.1       & 0.4 \\
                3& 50      & 18 & 0.68 & 0.5 & 2.1       & 1.0 \\
                4& 50      & 18 & 0.68 & 1.0 & 2.1       & 2.1 \\
                5& 50      & 18 & 0.68 & 2.0 & 2.1       & 4.1 \\
\hline
                6& 100      & 30 & 0.86 & 0.1 & 3.7       & 0.4 \\
                7& 100      & 30 & 0.86 & 0.2 & 3.7       & 0.7 \\
                8& 100      & 30 & 0.86 & 0.5 & 3.7       & 1.9 \\
                9& 100      & 30 & 0.86 & 1.0 & 3.7       & 3.7 \\
               10& 100      & 30 & 0.86 & 2.0 & 3.7       & 7.5 \\
\hline
               11& 200      & 45 & 1.06 & 0.1 & 6.7        & 0.7 \\
               12& 200      & 45 & 1.06 & 0.2 & 6.7        & 1.3 \\
               13& 200      & 45 & 1.06 & 0.5 & 6.7        & 3.4 \\
               14& 200      & 45 & 1.06 & 1.0 & 6.7        & 6.7 \\
               15& 200      & 45 & 1.06 & 2.0 & 6.7        &13.5 \\
\hline
\hline
        \end{tabular}
\end{table}

In order to study a fully-developed reaction wave, a planar wave $c(\boldsymbol{x},t=0)=c_L(\xi)$ was initially ($t=0$) released at $x^0=L_x/2$ such that
$\int_{-\infty}^{0} c_L(\xi) \mathrm{d} \xi =\int_{0}^{\infty} [1-c_L(\xi)] \mathrm{d} \xi$ and $\xi=x-x^0$, where, $c_L(\xi)$ is the pre-computed laminar-wave profile. Subsequently, evolution of this field $c(\boldsymbol{x},t)$ was simulated by solving Eq.~(\ref{A1}).
To enable periodic propagation of $c$ field along $x$-direction, the field is extrapolated outside the axial boundaries of the computational domain at each time step $t^n$ as follows;  $c(x',y,z,t^n)=c(x,z,t^n)$, where $x'=x+I \textrm{L}_x$ and $I$ is an arbitrary (positive or negative) integer number.
Consequently, Eq.~(\ref{A1}) is solved in the interval $x' \in [ \zeta (t^n)-\Delta, \zeta (t^n)+\Delta]$, where $\zeta (t^n)$ is the mean coordinate of a reaction wave on the $x'$-axis and $\Delta=0.45\textrm{L}_x$ in order to avoid numerical artifacts in the vicinity of $x'=\zeta (t^n) \pm 0.5\textrm{L}_x$.
In two remaining regions, i.e. $x'\in [\zeta(t^n)-0.5\textrm{L}_x, \zeta(t^n)-\Delta]$ and $x'\in [\zeta (t^n)+\Delta, \zeta(t^n)+0.5\textrm{L}_x]$, the scalar $c(t^{n})$ is set equal to zero (fresh reactants) and unity (products), respectively, because the entire flame brush is always kept within the interval of $x' \in [ \zeta (t^n)-\Delta, \zeta (t^n)+\Delta]$ in the present simulations. Finally, the obtained solution $c(x',y,z,t^{n})$ is translated back to the $x$-coordinate (see for details, \cite{PRE17RA,PoF17RA}).

Three turbulent fields were generated by specifying three different initial turbulent Reynolds numbers ${\rm Re}=50$, 100, and 200, which were increased by increasing the domain size $L$.
The increase in $L$ resulted in increasing
the longitudinal integral length scale $\ell_{11}$,
the Taylor length scale $\lambda=\sqrt{15 \nu u_0^2/\overline{\langle \varepsilon \rangle}}$, the Taylor scale Reynolds number ${\rm Re}_{\lambda} = u_0 \lambda/\nu$,
the turbulent time scale $\tau_{11}=\ell_{11}/u_0$,
and, hence, the Damk\"ohler number ${\rm Da_{_{\rm DNS}}}=\tau_{11}/\tau_{_F}$.
Here, $\overline{\langle \varepsilon \rangle}$ is the dissipation rate averaged over volume (angle brackets) and time at $t>5\tau_{\rm f}$ (overbars).
The simulation parameters are shown in Table I.
Because a reaction wave does not affect turbulence in the case of constant density $\rho$ and viscosity $\nu$, the flow statistics
were the same in all cases that had different $S_{_L}$, but the same ${\rm Re}$.
It is worth noting that the
longitudinal integral length scale
$\ell_{11}$ reported in Table I and used to evaluate
${\rm Da_{_{\rm DNS}}}$ was averaged over the
computational domain and time at $t>5 \tau_{\rm f}$ and was lower than its initial value $\ell_{\rm f}=L/4$.


When the width $L$ was increased by a factor of two, the numbers $N_x$, $N_y=N_x/4$, and $N_z=N_x/4$ were also increased by a factor
of two, i.e. $N_x=256$, 512, or 1024 at ${\rm Re}=50$, 100, or 200, respectively. Accordingly,
in all cases, the Kolmogorov length scale $\eta=(\nu^3/\overline{\langle \varepsilon \rangle})^{1/4}$
was of the order of the grid cell size $\Delta x$,
see Table I,
thus, indicating sufficient grid resolution.
Capability of the used grids for well resolving not only the Kolmogorov eddies, but also the reaction wave was confirmed
in separate (i) 1D simulations of planar laminar reaction waves
and (ii) 2D simulations \cite{YBB15} of laminar flames subject to
hydrodynamic instability \cite{LL}.
Moreover, the resolution
of the present DNS was validated by running simulations with the grid cell size $\Delta x$ decreased by
a factor of four at ${\rm Re}=50$, i.e. by setting $N_x$ equal to 1028.

In the next section, we will report the mean quantities $\overline{q}$ averaged over a transverse plane of $x=$const
and time at $5\tau_{\rm f} < t < t_{end}$, with $t_{end}$ being equal to $50 \tau_{\rm f}$ or even longer.
Moreover, we will present correlations between fluctuating quantities
$q'(t,{\bm x})=q(t,{\bm x})-\overline{q}(x)$.
Furthermore, using the computed dependencies of $\overline{c}(x)$, the dependencies of other mean quantities and
correlations on distance $x$ will be transformed to dependencies of these variables and correlations, respectively, on the mean
reaction progress variable $\overline{c}$.

\section{Results and discussion}

\begin{figure}
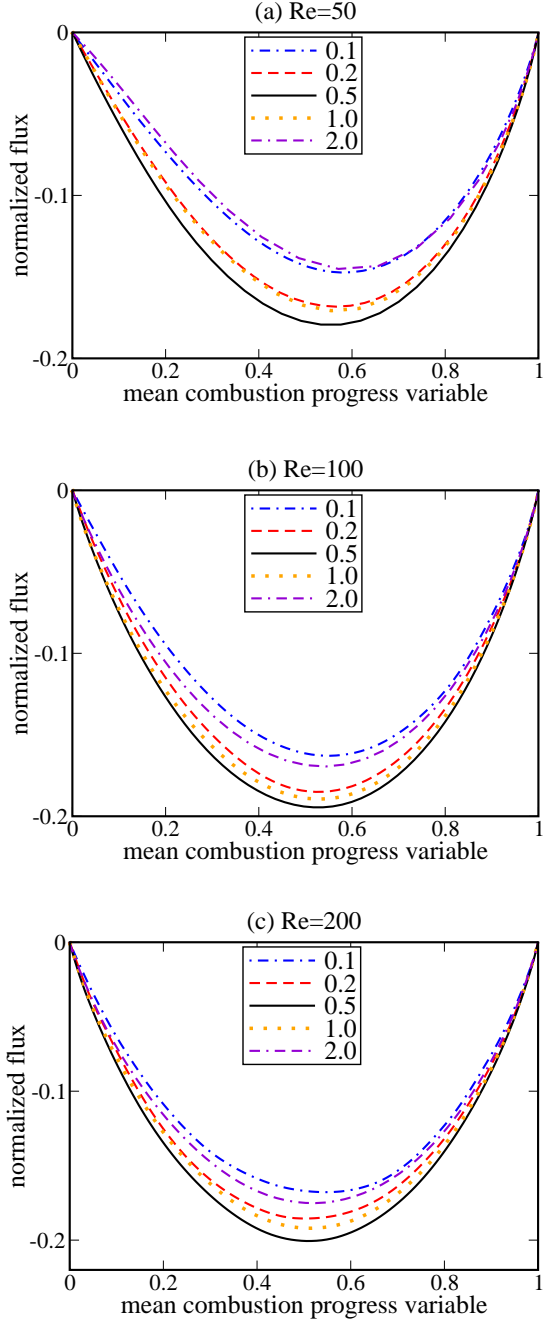

\vspace*{1mm} \centering
\includegraphics[width=8.0cm]{PRE_Fig1a.eps}
\includegraphics[width=8.0cm]{PRE_Fig1b.eps}
\includegraphics[width=8.0cm]{PRE_Fig1c.eps}
\caption{\label{Fig1} Dependencies of the normalized turbulent scalar flux $\overline{u' c'}/u_0$ on the mean reaction progress variable $\overline{c}$, computed
at different ratios of $S_{_L}/u_0$ specified in legends
for (a) ${\rm Re}=50$, (b) ${\rm Re}=100$, and (c) ${\rm Re}=200$.}
\end{figure}

\begin{figure}
\vspace*{1mm} \centering
\includegraphics[width=7.8cm]{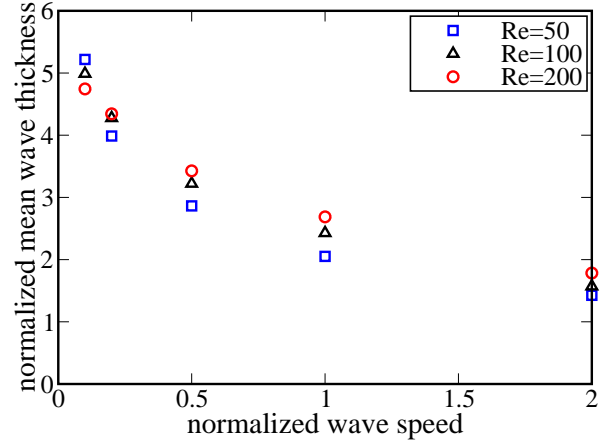}
\caption{\label{Fig2} Dependencies of the normalized mean turbulent wave thickness
$\delta_{\rm t}/\ell_{11}$ 
on the normalized wave speed
$S_{_L}/u_0$ computed at three different turbulent Reynolds numbers, ${\rm Re}$, specified in legends.
}
\end{figure}

\begin{figure}
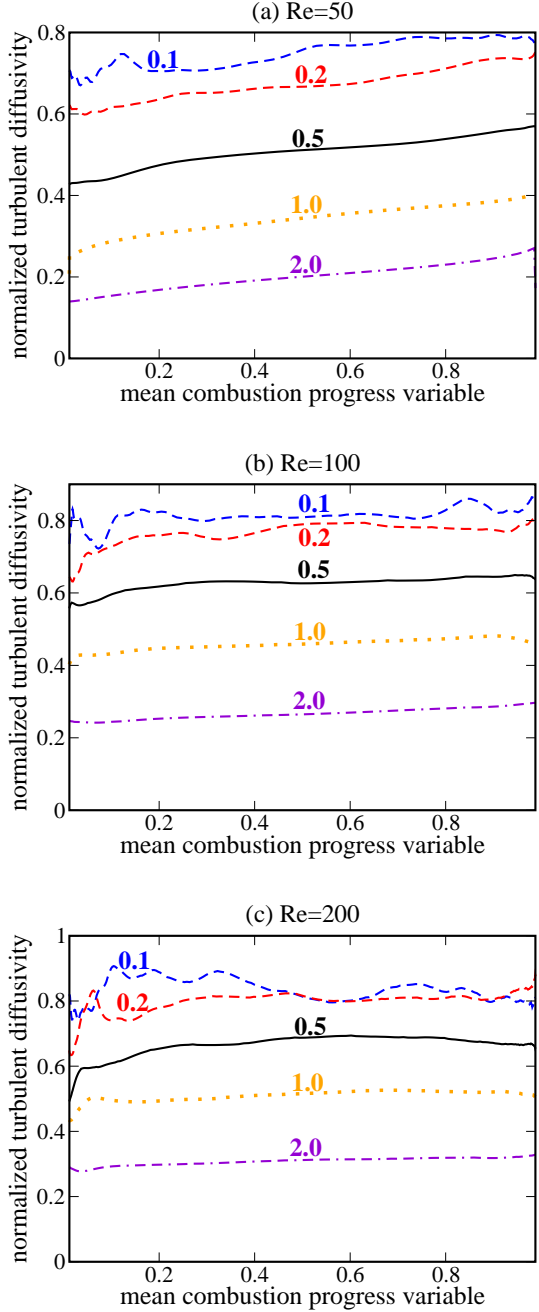

\vspace*{1mm} \centering
\includegraphics[width=8.0cm]{PRE_Fig3a.eps}
\includegraphics[width=8.0cm]{PRE_Fig3b.eps}
\includegraphics[width=8.0cm]{PRE_Fig3c.eps}
\caption{\label{Fig3}
Dependencies of the normalized
turbulent scalar diffusivity $D_T/(u_0 \ell_{11})$
on the mean reaction progress variable $\overline{c}$, computed for (a) ${\rm Re}=50$, (b) ${\rm Re}=100$, and (c) ${\rm Re}=200$.
Values of $S_{_L}/u_0$ are specified near curves.
}
\end{figure}

\begin{figure}
\centering
\includegraphics[width=8.5cm]{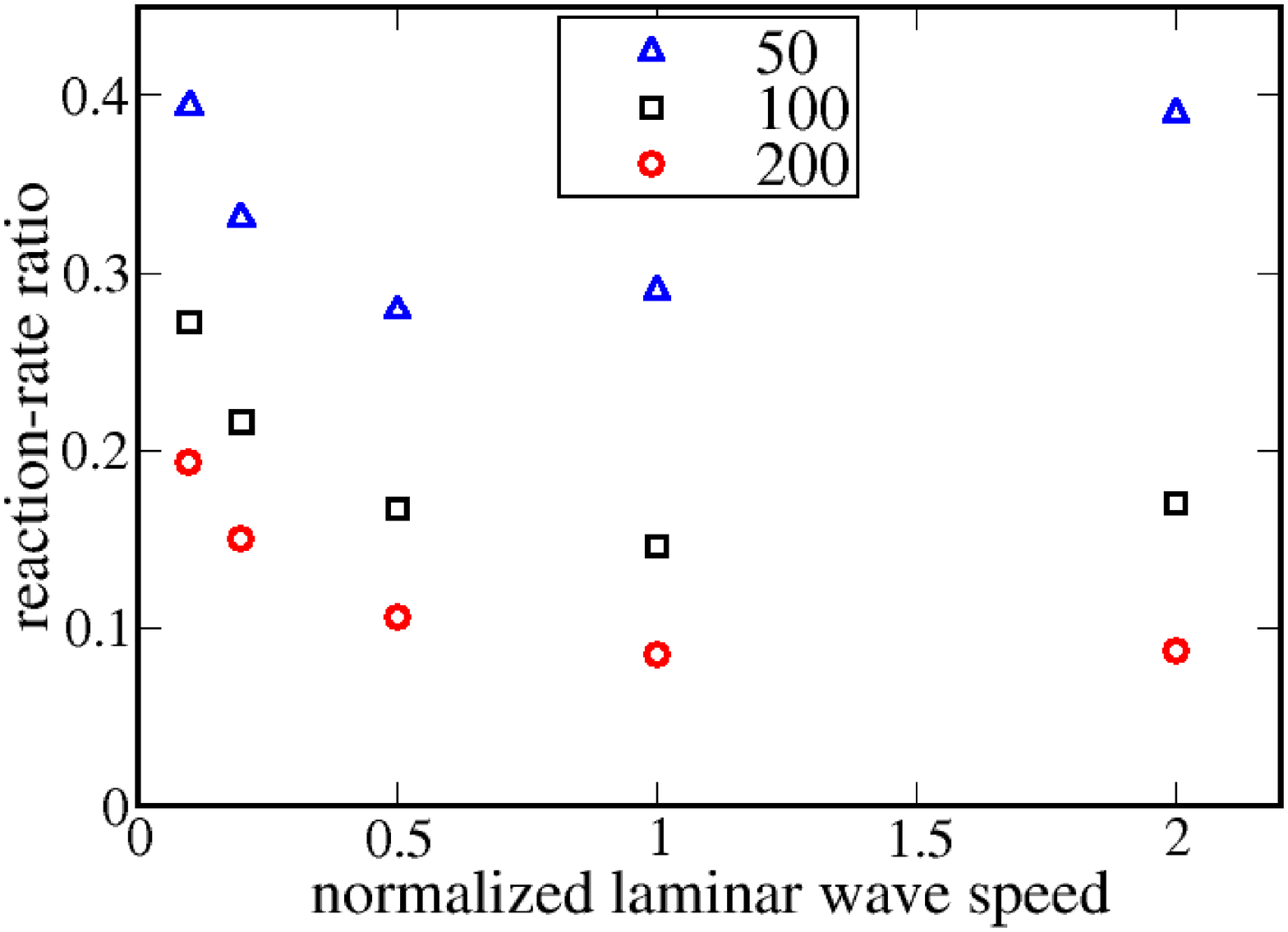}
\caption{\label{Fig4}
Reaction-rate ratio $\theta={\rm max}\{\overline{W}(\overline{c})\} /
{\rm max}\{W(c)\}$ vs. $S_L/u_0$. Symbols show DNS data, with the Reynolds numbers
Re being specified in legends.
}
\end{figure}


Figure~\ref{Fig1} shows dependencies of the normalized turbulent scalar flux $\overline{u' c'}/u_0$ versus the mean reaction progress variable $\overline{c}$, computed for (a) ${\rm Re}=50$, (b) ${\rm Re}=100$, and (c) ${\rm Re}=200$.
In unburnt or burnt mixture, the instantaneous progress variable is constant, $c=0$ or $c=1$, respectively.
This implies that in the two regions turbulent flux $\overline{c' {\bm u}'} =0$.
Inside the mean reaction wave the mean progress variable $\overline{c}$ varies
between 0 and 1. In this region the gradient ${\bm \nabla} \overline{c}$ does not vanish.
Since ${\bm \nabla} \overline{c}$ is positive in this region
(in the coordinate framework used in the paper),
the turbulent flux $\overline{c' {\bm u}'} = - D_T {\bm \nabla} \overline{c}$
is negative inside the mean reaction wave.
The absolute value of the turbulent flux $|\overline{c' {\bm u}'}|$ reaches maximum at the point where the gradient ${\bm \nabla} \overline{c}$ is maximum.
If the probability of deviation of the reaction wave from its mean position is described by the Gaussian distribution, the gradient
${\bm \nabla} \overline{c}$ is maximum at $\overline{c} =0.5$.
For instance, in various premixed turbulent flames, ${\bm \nabla} \overline{c}$ does peak at $\overline{c} =0.5$,
e.g. see Fig. 4.22 and Eqs. (4.34) and (4.35) in \cite{CRC}.
While, the flux magnitude depends on $S_{_L}/u_0$ and, hence, on ${\rm Da}_{_{\rm DNS}}$, see Table I, such variations in the flux
magnitude are sufficiently weak and non-monotonic, with the peak magnitude being obtained at a medium $S_{_L}/u_0=0.5$.

On the contrary, the mean turbulent wave thickness
$\delta_{\rm t}$ defined using the maximum gradient method, i.e.,
\begin{eqnarray}
\delta_{\rm t}
= \frac{1}{\max{ \{ \nabla_x \overline{c} \} }},
 \label{Eth}
\end{eqnarray}
decreases rapidly with the increase of the normalized wave speed
$S_{_L}/u_0$ and, hence, ${\rm Da}_{_{\rm DNS}}$,
see Fig.~\ref{Fig2}.
This numerical result is fully consistent with the theory, which predicts a decrease in $D_T$ with increasing Damk\"ohler number.
Under the DNS conditions, an increase in $S_L/u_0$ results in increasing ${\rm Da}$
and, therefore, decreasing $D_T$.
Consequently, $\delta_{\rm t} \propto \left[D_T({\rm Da}) \, \tau_c\right]^{1/2}$ decreases with increasing $S_L/u_0$.

\begin{figure}
\centering
\includegraphics[width=9.5cm]{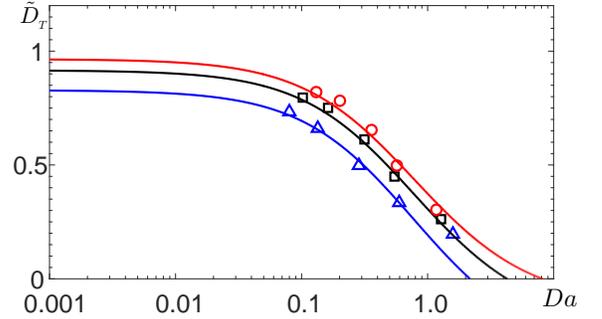}
\caption{\label{Fig5}
Theoretical dependence $\tilde D_T \equiv D_T/D^T_0$ versus Damk\"ohler number ${\rm Da}$ determined by Eq.~(\ref{CA17}) for different values of the Reynolds number ${\rm Re}=$ 50 (blue), 100 (black), 200 (red) at ${\rm Pr}=1$.
The DNS data on $\langle D_T \rangle$ normalized using
$u_0 l_{11}$ are shown in blue triangles (${\rm Re}=$ 50), black squares (${\rm Re}=$ 100), and red circles (${\rm Re}=$ 200).
}
\end{figure}

Accordingly, the gradient of the mean reaction
progress variable is increased by $S_{_L}/u_0$ (or ${\rm Da}_{_{\rm DNS}}$), whereas
turbulent diffusivity evaluated as follows,
\begin{eqnarray}
D_T(\overline{c}) = - \frac{\overline{u' c'}}{\nabla_x \overline{c}} ,
 \label{EDt}
\end{eqnarray}
is decreased with increasing $S_{_L}/u_0$ and ${\rm Da}_{_{\rm DNS}}$, see Fig.~\ref{Fig3}.
The decrease of the turbulent diffusion coefficient, $D_T$,
with the increase of the Damk\"ohler number observed in DNS,
agrees well with the developed theory.

Moreover, Fig.~\ref{Fig3} indicates that $D_T$ evaluated using Eq.~(\ref{EDt}) depends weakly on $\overline{c}$, thus, implying that the influence of the reaction on the turbulent diffusion coefficient may be characterized with a single mean turbulent diffusivity defined as follows
\begin{eqnarray}
\overline{\langle D_T \rangle} = \int_0^1 D_T(\xi) d \xi.
 \label{EmDt}
\end{eqnarray}

To compare values of the mean turbulent diffusion coefficient,
$\overline{\langle D_T \rangle}$, obtained in the simulations
with the theoretical predictions for $D_T({\rm Da})$
we need to take into account that the Damk\"ohler number,
${\rm Da}_{_{\rm DNS}}$, used in DNS is different from the Damk\"ohler number,
${\rm Da}$, used in the theory.
In the DNS, due to strong fluctuations in the scalar field $c$ and, especially, $W(c)$,
the mean reaction rate is characterized by a significantly larger chemical
time scale $\overline{\tau_c}$  when compared to the time scale $\tau_F$ associated
with the laminar $W(c)$.
A ratio of these two time scales $\theta=\tau_F/\overline{\tau_c}={\rm Da}
/{\rm Da}_{_{\rm DNS}}$ can be estimated as $\theta={\rm max}\{\overline{W}(\overline{c})\} /
{\rm max}\{W(c)\}$. The reaction-rate ratio $\theta$ versus $S_L/u_0$
is shown in Fig.~4 for different values of the Reynolds number Re.

Using the values of $\theta$ obtained in the DNS and plotted in Fig.~4,
we relate the Damk\"ohler number,
${\rm Da}$, used in the theory with ${\rm Da}_{_{\rm DNS}}$, used in DNS:
${\rm Da}= \theta \, {\rm Da}_{_{\rm DNS}}$.
In Fig.~\ref{Fig5} the mean turbulent diffusion coefficient, $\overline{\langle D_T \rangle}$, versus ${\rm Da}= \theta \, {\rm Da}_{_{\rm DNS}}$ obtained in the simulations (symbols) is compared with the theoretical predictions for $D_T$
given by Eq.~(\ref{CA17}).
Figure~\ref{Fig5} demonstrates very good agreement between results of DNS and
theoretical predictions.

\section{Conclusion}

The theory of turbulent diffusion in reacting flows
previously developed in \cite{EKR98,EKLR14},
has been generalized for
finite Reynolds numbers and the dependence of turbulent
diffusion coefficient versus two parameters, the Reynolds number
and Damk\"ohler number has been obtained.
Validation of the generalized theory of the effect of chemical reaction on turbulent diffusion using three-dimensional DNS of a finite thickness reaction wave propagation in forced, homogeneous, isotropic, and incompressible turbulence
for the first-order chemical reactions,
has revealed a very good quantitative
agreement between the theoretical predictions and
the DNS results.

\medskip

\begin{acknowledgements}
This research was supported in part by the Israel
Science Foundation governed by the Israeli
Academy of Sciences (Grant No. 1210/15, TE, NK, IR),
the Research Council of Norway under the FRINATEK (Grant 231444, NK, ML, IR),
the Swedish Research Council (RY), the Chalmers Combustion Engine Research Center (CERC) and Chalmers Transport and Energy Areas of Advance (AL),
State Key Laboratory of Explosion Science and Technology,
Beijing Institute of Technology (Grant KFJJ17-08M, ML).
This research was initiated during Nordita Program {\it Physics of Turbulent Combustion}, Stockholm (September 2016).
The computations were performed on resources provided by the Swedish National Infrastructure for Computing (SNIC) at Beskow-PDC Center.

\end{acknowledgements}

\end{document}